\documentstyle[referee]{mn}

\footnotesize
\newdimen\digitwidth    
\setbox0=\hbox{\rm0}
\digitwidth=\wd0
\catcode`!=\active
\def!{\kern\digitwidth}
\normalsize
  
\def\m4{Mark~IV}
\def\sma{\sigma_{\tau,A}}
\def\smb{\sigma_{\tau,B}}
 
\title[Baseband Recorder for Pulsar Observations]
{A Baseband Recorder for Radio Pulsar Observations}
\author[I. H. Stairs et al.]
{I. H. Stairs\footnotemark[1],
E. M. Splaver,
S. E. Thorsett\footnotemark[2],
D. J. Nice,
J. H. Taylor\\
Joseph Henry Laboratories and Physics Department,
       Princeton University, Princeton, NJ 08544 USA}
 
\date{\today}

\begin{document}

\maketitle
\newcommand{\setthebls}{
}
\footnotetext[1] {Present address: University of Manchester, Jodrell Bank Observatory, Macclesfield, Cheshire SK11 9DL; is@\-jb.\-man.\-ac.\-uk}
\footnotetext[2] {Present address: Department of Astronomy and Astrophysics, University of California, 1156 High St., Santa Cruz, CA 95064 USA}

\setthebls

\begin{abstract}

Digital signal recorders are becoming widely used in several subfields
of centimetre-wavelength radio astronomy.  We review the benefits and
design considerations of such systems and describe the Princeton
Mark~IV instrument, an implementation designed for
coherent-dedispersion pulsar observations.  Features of this
instrument include corrections for the distortions caused by coarse
quantization of the incoming signal, as well algorithms which
effectively excise both narrowband and broadband radio-frequency
interference.  Observations at 430\,MHz using the Mark~IV system in
parallel with a system using a 250\,kHz filter bank and incoherent
dedispersion demonstrated timing precision improvement by a factor of
3 or better for typical millisecond pulsars.

\end{abstract} 

\begin{keywords}
instrumentation:detectors -- instrumentation:polarimeters -- methods: observational -- pulsars: general -- pulsars: timing
\end{keywords}

\section{INTRODUCTION}\label{sec:intro}

A fundamental characteristic of radio astronomy is the use of
phase-coherent detectors, which coherently amplify the electromagnetic
radiation and preserve information about the phases of the wavefronts.
An ideal radio telescope backend should take advantage of this phase
coherency when recording and detecting data.  An effective way to do
so is to mix the telescope voltages to baseband, then Nyquist-sample
the data stream; however, the resulting data volumes quickly become
very large.  Historically, therefore, most wide-bandwidth
radio-astronomical observations have been carried out using analogue
detection and recording methods, typically a multi-channel
spectrometer.  An exception is Very Long Baseline Interferometry
(VLBI) observations, in which the voltages from different telescopes
are recorded to tape, then played back and combined on a custom
correlator to determine the visibility function.

In recent years, computer speeds have caught up with the data rates
needed for wide-bandwidth pre-detection sampling, to the extent that
computer clock rates are now within an order of magnitude of the
observing frequencies used in centimetre-wavelength astronomy.  These
continuing advances permit commercially-available hardware
to replace custom components in the construction of digital,
phase-preserving baseband recorders.  Such instruments are much more
flexible than hardware detection systems, permitting variable filters
and integration times, identification and excision of radio-frequency
interference, and, if the data are stored on tape, multiple
processing passes.  In this paper we describe the Princeton
\m4 system, a baseband recorder developed primarily for pulsar
astronomy and optimized for use with the Arecibo telescope; within the
next few years we expect similar instruments to become valuable in many
subfields of centimetre-wavelength astronomy, from radar ranging to
spectroscopy.

\section{DESIGN OF A BASEBAND RECORDER}\label{sec:hw}

One application of baseband recording is in high-precision timing and
polarimetric observations of millisecond pulsars.  Highly accurate
pulsar timing has applications not only in the study of the pulsars
themselves, but also in areas such as astrometry, time-keeping and
experimental tests of cosmology and general relativity.  An important
obstacle to high timing precision is dispersion of the pulses during
their propagation through the ionized interstellar medium.  This
phenomenon results in delays of lower-frequency radiation components
relative to the higher frequencies, and across a typical observing
bandwidth can amount to many (often hundreds or more) times the
intrinsic pulse widths.  The traditional pulsar timing and searching
instrument has been an analogue filterbank system, in which the bandpass
is subdivided into a number of channels, and the signal is detected in
each channel and shifted by the predicted dispersion delay in order
to align the pulse peak.  This method inevitably leaves residual
smearing within the channels, and the time resolution is limited to
the inverse of the channel bandwidth.  If, instead, the data are
sampled prior to detection, a frequency-domain ``chirp'' filter may be
applied to remove completely the predicted effects of dispersion and
align the pulse with no smearing.  Timing precision is
therefore greatly improved with this technique.

This ``coherent dedispersion'' method was pioneered more than two
decades ago \cite{hr75}, but until recently, the data storage and
processing limitations discussed above resulted in mostly narrowband,
hardware-based implementations, with special-purpose chips performing
the convolution of the data stream with the chirp function (e.g.,
Stinebring et al. 1992). \nocite{skn+92} While large, multi-channel
hardware dedispersion instruments are now in use (e.g., Backer et
al. 1997), \nocite{bdz+97} these systems record the data only after
convolution and detection and hence do not permit reprocessing or
interference excision.  Baseband recording coupled with software
dedispersion therefore offers a unique flexibility for the analysis of
pulsar signals, and various different implementations have been
presented in the literature \cite{jcpu97,wvd+98} and used for timing
and single-pulse studies.

The design considerations for a pre-detection digital recorder include
desired observing bandwidth, quantization resolution, recording
medium, processing capability and cost and availability of
components.  In continuum applications, a wide bandwidth is needed for
high signal-to-noise ratio; however the sampling and data rates scale
linearly with bandwidth.  To keep the data rate to a manageable level,
it is therefore necessary to accept coarsely-sampled data; some
techniques for optimizing signal quality in the case of 2-bit sampling
will be discussed below.  A further constraint on the feasible data
rate is the speed of the recording medium, typically hard disk or
magnetic tape.  Wider observing bandwidths also require more
computing power to process: the number of operations required may
increase more rapidly than linearly if, for instance, Fourier
Transforms are used in processing.  The optimal balance between the
different system components will depend on the goals determined for a
particular instrument.  For instance, the system described in Jenet
et al. (1997) uses a custom VLSI chip to provide 2-bit sampling
across 50\,MHz of bandwidth, and records data to a high-speed tape
recorder.  Data processing is then carried out on supercomputers.
Another implementation is the instrument described by Wietfeldt
et al.  (1998), in which a 16\,MHz bandpass is quantized at 2 bits,
and the data stream written to an adapted VLBI S2 recorder.  Upon
playback, the data may be analysed by a workstation or faster
computer.

The Princeton \m4 pulsar instrument was designed for use with the
8\,MHz-wide 430\,MHz line feed of the Arecibo radio telescope.  The
goal was to provide routinely usable baseband sampling and recording
across the full 8-MHz bandwidth using inexpensive hardware,
commercially available recording media and an affordable dedicated
processor.  It provides somewhat greater flexibility than the systems
discussed above, allowing 2-bit sampling across 10\,MHz or 4-bit
sampling across 5\,MHz of bandwidth.  Early prototypes of the system
have been discussed in Shrauner et al.  (1996) and Shrauner
(1997).  \nocite{ssd+96,shr97} The final version, with a design based
on the second prototype, is currently installed at the upgraded
Arecibo Observatory.

\section{MARK IV HARDWARE AND SOFTWARE IMPLEMENTATION}

\begin{figure*}
\setlength{\unitlength}{1cm}
\begin{picture}(12,12)
\put(0.,-1){\includegraphics{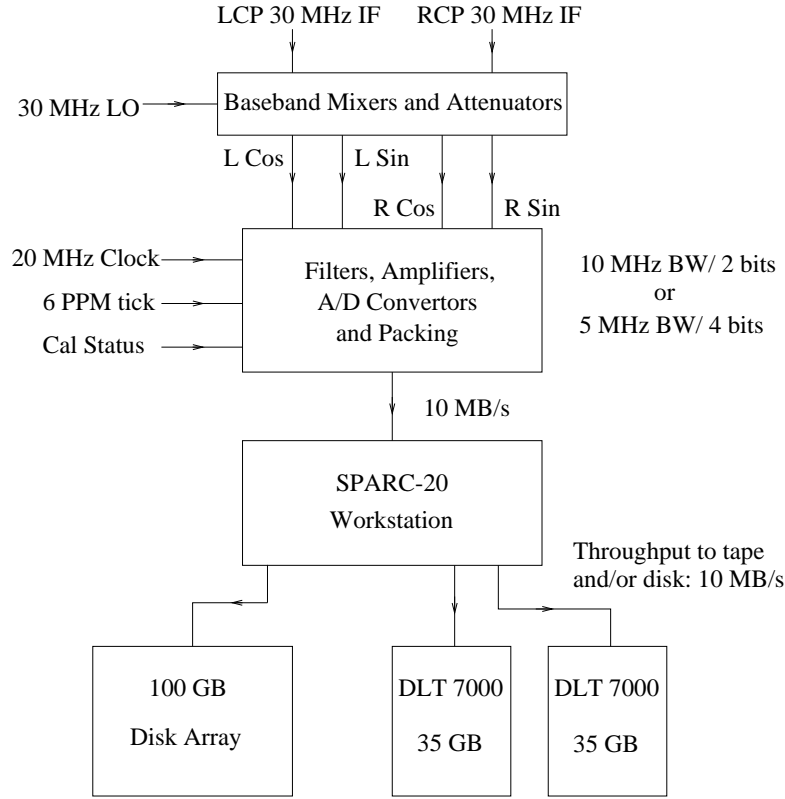}}
\end{picture}
\caption
[] {Block diagram of the \m4 hardware.}
    \label{fig:hwblock}
\end{figure*}

The \m4 instrument accepts intermediate-frequency (IF) signals of
bandwidth $B$ (either 5 or 10 MHz) centred at 30\,MHz.  Adjustable
attenuators regulate the signal strength, and the voltages are then
mixed to baseband with quadrature local oscillators (LOs) at 30 MHz,
producing, for each of two orthogonal polarizations, a real and an
imaginary signal with passband 0 to $B/2$.  These four signals are
low-pass-filtered with a suppression in excess of 60 dB at $B/2$, then
4-bit digitized at rate $B$, in accordance with the Nyquist theorem.
Shift registers and multiplexers pack the samples such that all 4 bits
are retained for the 5~MHz bandpass, while only the 2 most significant
are kept for the 10~MHz bandpass.  Thus the overall data rate is
10\,MB/s regardless of bandwidth.  This flow of data is piped through
a DMA card into a SPARC-20 workstation and then onto a combination of
hard disks and Digital Linear Tapes (DLTs) for off-line processing.  The
digitizer/packer board is clocked by a 20~MHz signal which is tied to
the observatory time standard.  The data timestamp is generated by a
10-second tick tied to the same external clock.  The status of an
injected noise signal may be monitored for later use in calibration.
A block diagram of the system is shown in Figure~\ref{fig:hwblock}.

\begin{figure*}
\setlength{\unitlength}{1cm}
\begin{picture}(12,8)
\put(0.,-4){\includegraphics{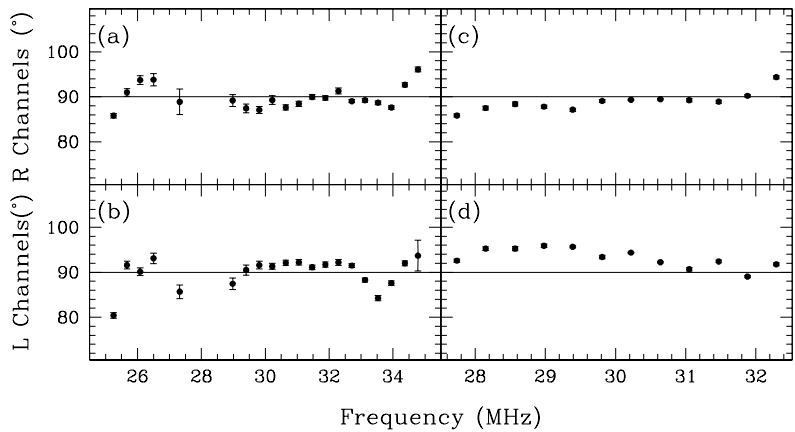}}
\end{picture}
\caption[]{
Mean phase differences at harmonics of
  414\,kHz between real and imaginary components of the left and right
  channels.  Panels (a) and (b) display results for the 10\,MHz
  bandpass; panels (c) and (d) for the 5\,MHz bandpass.  Some of the
  harmonics present in the 5\,MHz bandpass do not appear in the
  10\,MHz plots because the 2-bit quantization was not sensitive to
  the lower signal strength.} 
    \label{fig:phasediff}
\end{figure*}

In such a recording system hardware-related systematic errors may
arise from the baseband mixers, from the low-pass filters, or from the
amplifiers and attenuators in the signal path; care has been taken to
minimize these errors.  To verify the orthogonality of the baseband
quadrature signals, a comb of harmonics of 414\,kHz was used as an
input signal.  The data streams from the four input channels were
separately Fourier-transformed and the phase differences between the
real and imaginary components of the left and right channels were
calculated at each harmonic.  The results are plotted in
Figure~\ref{fig:phasediff}.  The means cluster around $90^\circ$, as
they should, with only slight variations across either bandpass.  The
absence of some of the harmonics in the 10\,MHz bandwidth plots
(panels (a) and (b)) is due to the low intrinsic amplitude of the test
signal at these frequencies.

The filters were selected for flat amplitude response with frequency;
for other applications, constant response in phase may be more
appropriate.  Measurements indicate that the total phase rotation
across the bandpass of one of the 5-MHz filters amounts to some 2.5
turns of phase.  This extra shift could be incorporated into the chirp
function; however, it is negligible relative to the thousands of turns
of phase typically induced by dispersion.  The amplitude response of
the entire system is nearly flat as a function of frequency out to the
knee of the filters.

\subsection{Signal processing}\label{sec:proc}

The \m4 system produces packed data at a rate of 10\,MB/s, or
35\,GB/hr.  Analysis of this data stream in real time would require
8-10~Gflops.  As an affordable alternative to a supercomputer, we use
a 1.25~Gflop parallel processor optimized for Fast Fourier Transforms
(FFTs), which form the core of the analysis.  This machine, the
SAM-350 from Texas Memory Systems, Inc., consists of 512\,MB of fast
memory, a DEC Alpha AX27 scalar processor, a parallel-processor
board containing customized chips and an additional processor to
handle communications.  The fast memory may be accessed by the AX27,
the parallel-processor board or a host workstation via an SBUS card.

Modeling the interstellar medium as a tenuous electron plasma permits
the calculation of the ``chirp'' function used in the dedispersion
analysis \cite{hr75}:
\begin{equation}
\label{eqn:Hf}
H(f_0+f_1) \,=\, {\rm exp} \left[ 2\pi i \,\frac{{\rm
DM}}{2.41\times10^{-10}}\,\frac{f_1^2}{f_0^2(f_0+f_1)} \right],
\end{equation}
where $f_0$ is the central observing frequency in MHz, $|f_1| \leq
B/2$, where $B$ is the observing bandwidth, and the dispersion
measure, given by ${\rm DM} = \int_0^dN_edz$, is the integrated
electron density along the line of sight to the pulsar, measured in
pc\,cm$^{-3}$.  Coherent dedispersion is performed by transforming a
segment of the baseband data to the Fourier domain, multiplying by the
inverse of this chirp function and then transforming back to the time
domain, with suitable overlap of successive data segments.  In
practice, the inverse FFT is taken in 2, 4 or 8 parts, splitting the
bandpass into as many sub-bands.  This permits the monitoring of
potentially variable data quality (perhaps due to interference or
scintillation) in different parts of the band, as well as the Faraday
rotation of the linear polarization across the band.

Four cross-products are formed from the dedispersed data stream:
$|L|^2$, $|R|^2$, Re($L^*R$) and Im($L^*R$).  These detected time
series may be recorded directly and used in the analysis of single
pulses.  Usually, however, the data points for each sub-band and each
cross-product are folded modulo the predicted pulse period and are
summed into 10-second accumulated pulse profiles.  The folded products
are calibrated using the observed magnitude of a noise calibrator
which is pulsed for one minute after a pulsar observation; the
strength of the noise calibrator in Jy is known from comparisons with
catalogue flux calibration sources.  The Stokes parameters are readily
calculated from the four recorded products; parallactic angle
correction and polarimetry have been discussed elsewhere
\cite{stc99}.

Precise calculation of pulse times-of-arrival (TOAs) is essential to
pulsar timing.  The total-intensity folded profiles are
cross-correlated with a standard template to measure the phase offset
of the pulse within the profile \cite{tay92} The offset is added to
the time of the first sample of a pulse period near the middle of the
data set to yield an effective pulse time-of-arrival.

\subsection{Signal quality}\label{sec:quant}

Digital sampling is inherently a non-linear process: with less than an
infinite number of bits, noise will inevitably be added to the signal.
As the determination of precise times-of-arrival depends on accurate
pulse profile shapes, preservation of the pulse shape is the primary
concern in a coarsely-quantized pulsar observing system.  The extreme
case of 1-bit sampling, in which ``on'' and ``off'' values are
assigned by comparing each sample to a running mean, yields the
noisiest reproduced signal.  Data rate considerations have motivated
the choice of 2- and 4-bit sampling for the \m4 system, an improvement
on 1-bit sampling, but still coarse.  This quantization will
necessarily affect the observed pulse shapes and signal-to-noise
ratios in statistically predictable ways.

Preserving the pulse profile in the case of 2-bit quantization
requires some care.  In this quantization process, the decision
thresholds are $-v_0$, 0 and $+v_0$ and the values assigned to the
output levels are $-n$, $-1$, 1 and +$n$, where $n$ is not
necessarily an integer.  For a randomly fluctuating signal, Cooper
(1970) \nocite{coo70} finds a recoverable signal-to-noise ratio of
0.88 using $n = 3$ and $v_0$ equal to the root-mean-square voltage of
the input signal, $v_{\rm rms}$.  However, in pulsar observations
there is an additional complication: the signal is dedispersed after
quantization.  For the straightforward choice of $v_0 = 1.0v_{\rm
rms}$ and $n = 3$, this process can result in the appearance of dips
to either side of the pulse when power is shifted to the aligned peak
from the neighbouring regions of the profile \cite{ja98}.  This
effect is most pronounced when the dispersion time across the
observing bandwidth is comparable to the pulse width.

\begin{figure*}
\setlength{\unitlength}{1cm}
\begin{picture}(12,12)
\put(0.,-2){\includegraphics{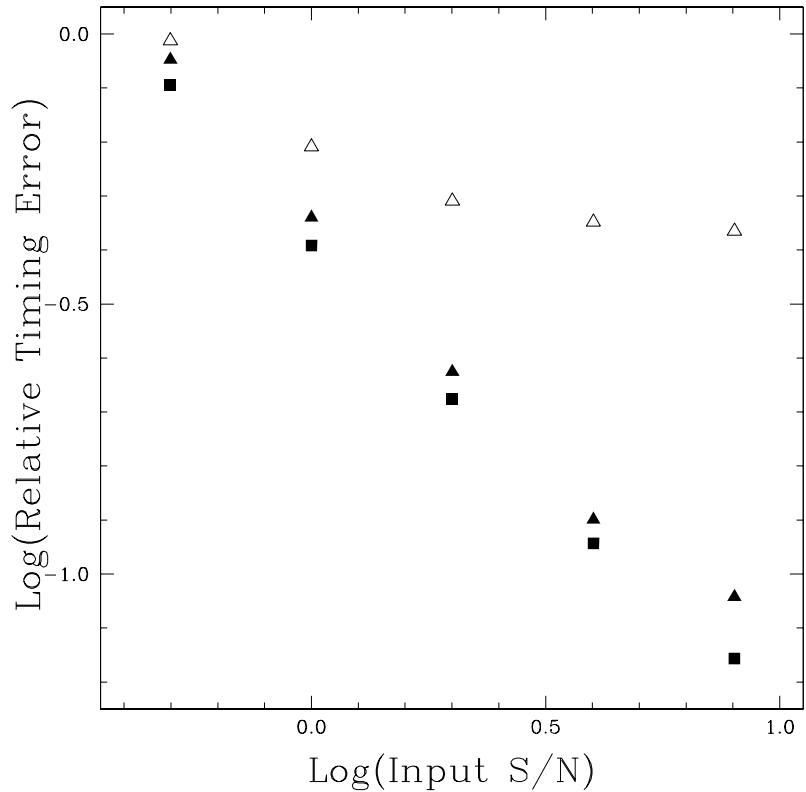}}
\end{picture}
\caption[]{
    Root-mean-square uncertainties in the offset between the pulse
    recovered from the Monte Carlo quantization simulations and the
    original template for PSR~B1534$+$12, as determined by the pulsar
    timing cross-correlation routine.  The open triangles denote the
    results for 2-bit quantization, $v_0 = 1.0v_{\rm rms}$, $n = 3$,
    the solid triangles 2-bit quantization, $v_0 = 1.4v_{\rm rms}$, $n
    = 4$ and the solid squares 4-bit quantization, $v_0 = 0.59v_{\rm
    rms}$ and evenly-spaced output levels.  The latter two
    combinations show approximately linear behaviour with increasing
    input signal-to-noise ratio, indicating that the pulse profile is
    well-preserved under these quantization schemes.}\label{fig:1534}
\end{figure*}

To find the quantization levels which minimize dips, we performed
Monte Carlo simulations of observations with various quantization
thresholds.  The simulations used a high-resolution pulse profile of
PSR~B1534$+$12, dispersing, quantizing, dedispersing and accumulating
5000 of these pulses in each of 10 trials at each of several different
initial signal-to-noise levels.  The resulting profiles were then
cross-correlated against the original profile, following the standard
procedure used in pulsar timing.  The uncertainty in the
cross-correlation fit is therefore a measure of both the strength of
the reproduced profile and its resemblance to the original.
Figure~\ref{fig:1534} plots the root-mean-square cross-correlation
uncertainties against initial signal-to-noise ratio for 4-bit
quantization with $v_0 = 0.59v_{\rm rms}$ and evenly-spaced output
levels, for 2-bit quantization with $v_0 = 1.40v_{\rm rms}$ and $n =
4$ and for 2-bit quantization with $v_0 = 1.0v_{\rm rms}$ and $n = 3$.
It is apparent that the first two cases retain fairly good linearity
across the range in question, where the strength of the individual
pulses ranges from 0.5 of the system noise to 8 times the system
noise, whereas the third case yields very poor reproductions of the
original profile, particularly at higher signal-to-noise ratios.  As
the combination of $v_0 = 1.40v_{\rm rms}$ and $n = 4$ for 2-bit
quantization best eliminates dips and preserves the pulse shape, while
making the final signal-to-noise ratio roughly 0.82 of the undispersed
value, these parameters were adopted for all 2-bit observations.

\begin{figure*}
\setlength{\unitlength}{1cm}
\begin{picture}(12,12)
\put(0.,-2){\includegraphics{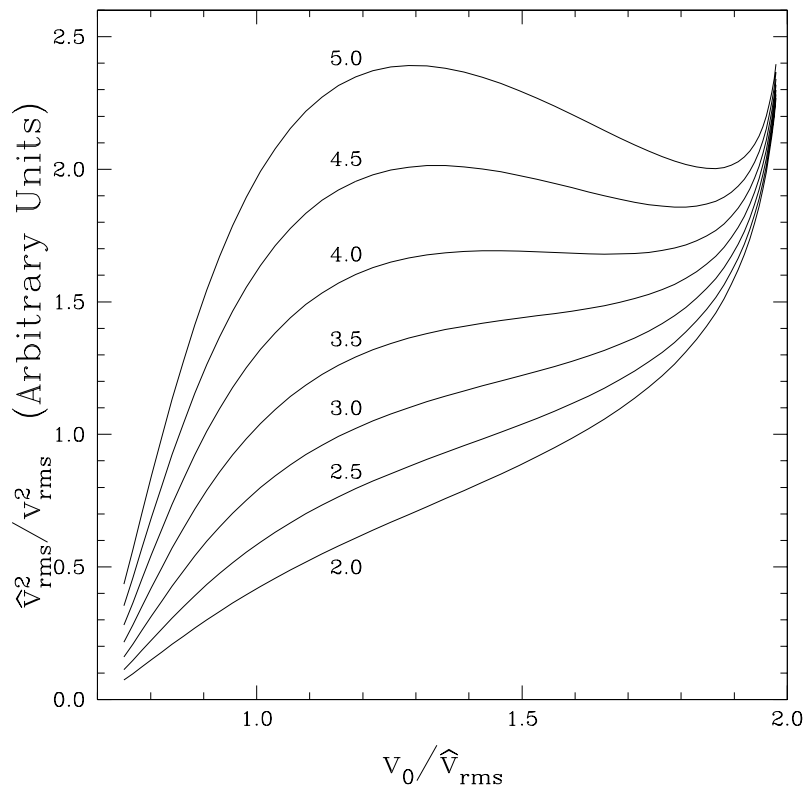}}
\end{picture}
\caption
[] {Ratio of quantized power to unquantized power (in arbitrary
    units) for different values of the quantization decision threshold
    $v_0$ and outer level spacing $n$.  Good linearity is achieved for
    $n = 4.0$ in the domain around $v_0 = 1.4\hat{v}_{\rm rms}$, where
    the response curve is nearly flat.}  \label{fig:quant}
\end{figure*}

In practice, the quantized $\hat{v}_{\rm rms}$ is estimated during
data acquisition by calculating histograms of the incoming data using
$n=3$.  The attenuation of the input voltage is then adjusted until
the measured $\hat{v}_{\rm rms}\,=\,0.71v_0$.  Thus the quantization
threshold is set based on the quantized power rather than the
unquantized power.  Assuming a gaussian distribution of counts,
$\hat{v}^2_{\rm rms}$ may be calculated analytically for any given
$v_0$ and $n$:
\begin{equation}\label{eqn:vrms}
\hat{v}^2_{\rm rms}\,=\, \frac{1}{4} \left(n^2 +
(1-n^2)P\left(0.5,0.5\left[\frac{v_0}{v_{\rm rms}}\right]^2\right)\right),
\end{equation}
where $P(a,x)$ is the incomplete gamma function and the factor of 1/4
is an arbitrary normalization.  Based on this calculation, the
combination of $n=4$ and an input $v_{\rm rms}$ voltage such that $v_0
\sim 1.4\hat{v}_{\rm rms}$ yields very good power linearity.  (Note
that for $n=3$, $v_0 = 1.4\hat{v}_{\rm rms}$ implies $v_0 \simeq
1.5v_{\rm rms}$.)  The linearity as a function of $v_0$ and $n$ may be
seen in Figure~\ref{fig:quant}.  The flat response for $ n = 4$ in the
region around $v_0 \sim 1.4\hat{v}_{\rm rms}$ confirms the conclusions
of the simulations.

Other procedures may still be necessary in order to recover the
correct pulse shape.  If there are very large power variations, for
instance, the quantized power levels will often underestimate the true
power and dips will result despite the $n=4$ correction.  Jenet and
Anderson (1998) \nocite{ja98} overcome this problem in their 50-MHz
baseband recorder by calculating the digitized $\hat{v}_{\rm rms}$
many times over the course of the predicted pulse period and
dynamically adjusting $n$ to preserve linearity.  However, this method
is not practical to compensate for very rapid fluctuations in system
temperature during observations of the fastest pulsars.  For example,
the power in the pulsar signal may cause the telescope system
temperature to fluctuate by a factor of 10 on time time scale of
1/1000 of a period, perhaps 2\,$\mu$s.  With a sample rate of 10\,MHz,
this would allow only 20 samples per noise-level calculation, leading
to estimation errors on $v_{\rm rms}$ on the order of $20^{-1/2}\sim
20\%$, too large to accurately adjust the output levels.  Jenet and
Anderson (1998) also discuss a correction for ``scattered power''
which becomes apparent in their band after dynamically setting $n$.
With the fixed $n=4$ scheme, scattered power does not appear to be a
problem.  It is possible that the level-setting happens to be
optimized to make the strength of residual dips and the scattered
power exactly equal; however, as pulse-shape distortions are not
evident, it appears that the existing scheme provides an adequate
balance between efficient performance and high-quality signal
reproduction.

\section{OBSERVATIONS WITH THE MARK IV INSTRUMENT}

The \m4 system and its prototypes have already been used for a variety
of different observing purposes, such as studying single and giant
pulses in PSR~B1937+21 \cite{cstt96}, polarimetry of a large number
of millisecond pulsars \cite{stc99}, high-precision timing of
millisecond pulsars, including the double--neutron-star binary
PSR~B1534+12 \cite{sac+98} and, using a software-synthesized
filterbank, searches for the extremely fast pulsars that may exist in
globular clusters.  There have also been preliminary observations
using the instrument as a backend for radar studies of asteroids; the
instrument is clearly proving to be as flexible as hoped.  Below we
address the issue of greatest concern in pulsar timing: the time
resolution obtainable with the new instrument relative to that
obtained with the filterbank system it supersedes.  We also discuss
the short-term timing stability of the instrument and some
interference-excision techniques developed during the course of the
first observations with the completed system.

\subsection{Comparison with analogue filterbank}\label{sec:m3m4}

Pre-detection coherent dedispersion, as implemented in Mark~IV, is
expected to allow marked improvement in the precision of pulsar timing
experiments relative to post-detection dedispersion systems,
particularly in cases with substantial dispersion smearing within
individual spectral channels.  To investigate this, a series of test
observations were made using the 305\,m Arecibo telescope at 430\,MHz.
The Mark~IV system collected data in parallel with the earlier
Mark~III system \cite{skn+92}, which was commonly used for pulsar
timing experiments before the recent Arecibo upgrade.  For these
observations, the Mark\,III system used a $2\times32\times250$\,kHz
analogue filter bank to detect signals across an 8\,MHz passband in two
polarizations.  The detected signals were low pass filtered with a
time constant of 100\,$\mu$s, after which they were digitized and
folded modulo the predicted topocentric pulse period.

\subsubsection{Expected precision of time of arrival measurements}

The measurement of a pulse arrival time is made by fitting an observed
pulse profile, $p(t)$, to a scaled, shifted high signal to noise ratio
standard profile, $s(t)$:
\begin{equation}
p(t) = a + b\,s(t-\tau)+g(t)
\end{equation}
where a, b and $\tau$ are constants, and $g(t)$ represents random
radiometer and background noise, and where $0<t<P$, with $P$ being the
pulsar period.  The quantity of greatest interest is the time shift
$\tau$ which, when added to the integration start time (along with a
mid-scan correction), gives the pulse time of arrival.

The uncertainty in the arrival time is dominated by the uncertainty in
$\tau$, $\sigma_\tau$.  In the limit where the pulsar is much weaker
than the system noise, this uncertainty is \cite{dr83}:
\begin{equation}
\sigma_\tau=\frac{\sigma_n/b}{\left[\int_0^P(s'(t))^2dt\right]^{1/2}},
\end{equation}
where $s'(t)$ is the time derivative of the standard profile and
$\sigma_n$ is a measure of system noise.  
According to the radiometer equation,
\begin{equation}
\sigma_n \sim 1/(Bt)^{1/2}
\end{equation}
where $B$ is the bandwidth of detected radiation and $t$ is the
integration time.  The ratio of timing precision of two observing
systems, A and B, can therefore be written
\begin{equation}
\frac{\sigma_{\tau,A}}{\sigma_{\tau,B}} = 
  \left(\beta \frac{B_B t_B}{B_A t_A}\right)^{1/2},
\end{equation}
where the shape factor $\beta$ is defined by
\begin{equation}
\beta = \frac{\int\,(s'_B(t))^2\,dt}{\int\,(s'_A(t))^2\,dt}.
\end{equation}

In the case of interest, system A is the Mark~IV coherent dedispersion
system and the profile $s_A(t)$ is a close representation of the
intrinsic pulse profile $s_{\rm int}(t)$ as emitted by the pulsar.
System B is the Mark~III post-detection dedispersion system.  The
observed pulse shape $s_B(t)$ is the intrinsic pulse profile convolved
with the effects of dispersion smearing and the detector time
constant, calculated as follows.

The effect of dispersion smearing may be quantified by noting that the
transmission function of the filters used by Mark~III are well
described by a Gaussian,
\begin{equation}
\phi(f_1) \sim \exp [ - (f_1)^2/w^2]
\end{equation}
where $f_0$ is the centre frequency of the sub-band defined by the
filter, $f_1$ is a frequency within the filter sub-band, and with
$w=125$\,kHz is the filter half-width.  To first order, the signal at
frequency $f\,=\,f_0+f_1$ from a pulsar with dispersion measure DM is
delayed by an amount $t-t_0=(1/\alpha)\,f_1$ relative to the centre of
the sub-band, where the dispersion slope is
\begin{equation}
\alpha = \frac{1.205\times10^{-4}\ f_0^3}{\rm DM}\ 
        \frac{{\rm pc\,cm}^{-3}}{\rm MHz^2\,s}.
\end{equation}
Thus the intrinsic pulse profile is convolved with
\begin{equation}
\phi(\alpha t) = \exp [- (\alpha\,t/w)^2].
\end{equation}
Low pass filtering of the detected signal has the effect of further
convolving the pulse profile with
\begin{equation}
\exp(-t/t_d),
\end{equation}
where $t_d=100\,\mu$s is the detector time constant. 

\begin{figure*}
\setlength{\unitlength}{1cm}
\begin{picture}(12,12)
\put(0.,0){\includegraphics{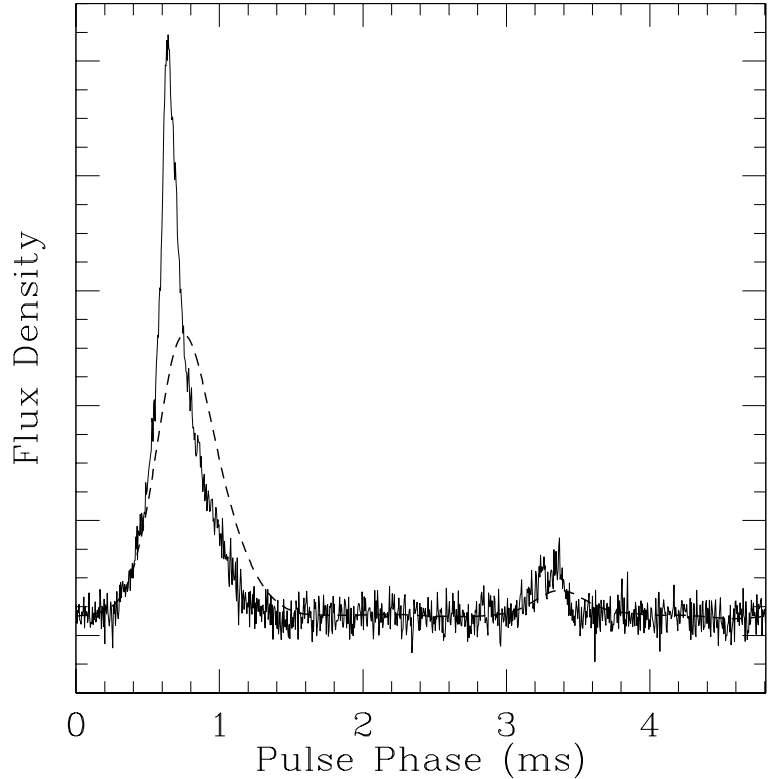}}
\end{picture}
\caption[]
{Pulse profile of PSR J2322+2057.  {\it Solid line:}
 pulse profile at 430\,MHz observed with Mark IV coherent
 dedispersion.  {\it Dashed line:} The same profile, filtered by
 incoherent dedispersion as described in the text.} \label{fig:prof}
\end{figure*}

The pulse profile expected to be observed by Mark~III, $s_B(t)$, can
be predicted by applying these convolutions to the intrinsic pulse
profile as measured by Mark~IV, $s_A(t)$.  An example of a Mark~IV
profile shape and the same shape filtered by the Mark~III system is
shown in Figure~\ref{fig:prof}.  From such profiles, the shape factor
$\beta$ and expected improvement in timing precision,
$\sigma_{\tau,A}/\sigma_{\tau,B}$ can be derived.  Values for the
sources in our test observations are given in Table~\ref{tab:resid}.

\subsubsection{Observations and results}

For these observations, pulsars were selected for which dispersion
smearing across a single channel was comparable to or larger than the
intrinsic pulse width and the filter bank time constant.  Sources were
observed simultaneously with the Mark~III and Mark~IV systems,
typically for 29 or 58 minutes on a given day.  Mark~III covered a
bandwidth of $B_A=8$\,MHz (except for one observation; see
Table~\ref{tab:resid}).  Mark~IV covered a bandwidth of $B_B=5$\,MHz
with 4-bit sampling.  Integration time for Mark~III varied; see
Table~\ref{tab:resid}.  Integration time for Mark~IV was fixed at
$t_A=190$\,s.  Combining these bandwidths and integration times
resulted in the predicted ratio of timing residuals (see
Table~\ref{tab:resid}).

\begin{table*}
\caption{Comparison of Mark~III and Mark~IV timing residuals\label{tab:resid}}
\vspace{2mm}

\begin{center}
\begin{tabular}{lrclrrcrrc}
\hline
\hline
\multicolumn{1}{c}{Pulsar}     &
\multicolumn{1}{c}{P}   &  
\multicolumn{1}{c}{DM}   & 
\multicolumn{1}{c}{Date }    & 
\multicolumn{1}{c}{$t_B\rm$} & 
\multicolumn{1}{c}{$\beta$} & 
\multicolumn{1}{c}{Predicted}    & 
\multicolumn{3}{c}{Observed}          \\
\multicolumn{1}{c}{Name}      &   
\multicolumn{1}{c}{(ms)} &
\multicolumn{1}{c}{(pc\,cm$^{-3}$)} &
\multicolumn{1}{c}{(MJD)} &
\multicolumn{1}{c}{(s)}       &  & 
\multicolumn{1}{c}{$\smb/\sma$}  & 
\multicolumn{1}{c}{$\sma$}      &  
\multicolumn{1}{c}{$\smb$} &  
\multicolumn{1}{c}{$\smb/\sma$}   \\
 & & & & & & & 
\multicolumn{1}{c}{($\mu$s)}     & 
\multicolumn{1}{c}{($\mu$s)} &   \\
                                                                               
\hline                      
J0621+1002 &     28.8 & 36.6 & 51138     &     123  & 0.56     &  0.56        & 5.0        &   2.8    &        0.55    \\
J0751+1807 &  3.5 & 30.3 & 51138     &     183  & 0.21     &  0.26        & 9.0        &   3.2    &        0.36    \\
J1022+1001 &     16.5 & 10.2 & 51138     &     123  & 0.62     &  0.62        & 3.2        &   1.5    &        0.48    \\    
J2019+2425 & 3.9 & 17.2 & 50995     &     130  & 0.30     &  0.31        &    27.3        &   3.5    &        0.13    \\
J2019+2425 &  3.9 & 17.2 & 51138     &     130  & 0.30     &  0.31        &    19.1        &   6.8    &        0.37    \\
J2322+2057 &  4.8 & 13.4 & 51138$^\ast$    &     183  & 0.33     &  0.38        & 4.5        &   1.3    &        0.30  \\
J2322+2057 &  4.8 & 13.4 & 51139     & 23  & 0.33     &  0.15        &    33.4        &   3.8    &        0.11    \\ \hline
\end{tabular}
\end{center}
\begin{raggedright}
Note. -- Mark IV: B=5\,MHz, t=190\,s always; Mark III: B=8\,MHz except
observation noted by *, for which B=6.75\,MHz.
\end{raggedright}
\end{table*}

Data from each observing system was reduced by calculating pulse
arrival times and fitting to a pulsar timing model with the standard
{\sc tempo} ({\tt http://pulsar.princeton.edu/tempo}) program.  The
root-mean-square values of the residuals from these fits are tabulated
as $\sigma_{\tau,A}$ and $\sigma_{\tau,B}$.

In most cases Mark~IV improved the timing precision as much as, or
more than, expected.  There are several possible sources of
improvement beyond the predicted values.  (1) Mark~IV has better
interference removal, since individual points in the time series are
examined, as described in \S\ref{sec:rfi} below. (2) Variations in
pulsar flux density within individual spectral channels of the
Mark~III system cause non-uniform dispersion smearing, resulting in
biased arrival times.  (3) Small errors in the calibration of the
analogue filter bank could lead to errors of several microseconds.

\subsection{Short-term timing stability}\label{sec:stable}

\begin{figure*}
\setlength{\unitlength}{1cm}
\begin{picture}(12,12)
\put(0.,-2){\includegraphics{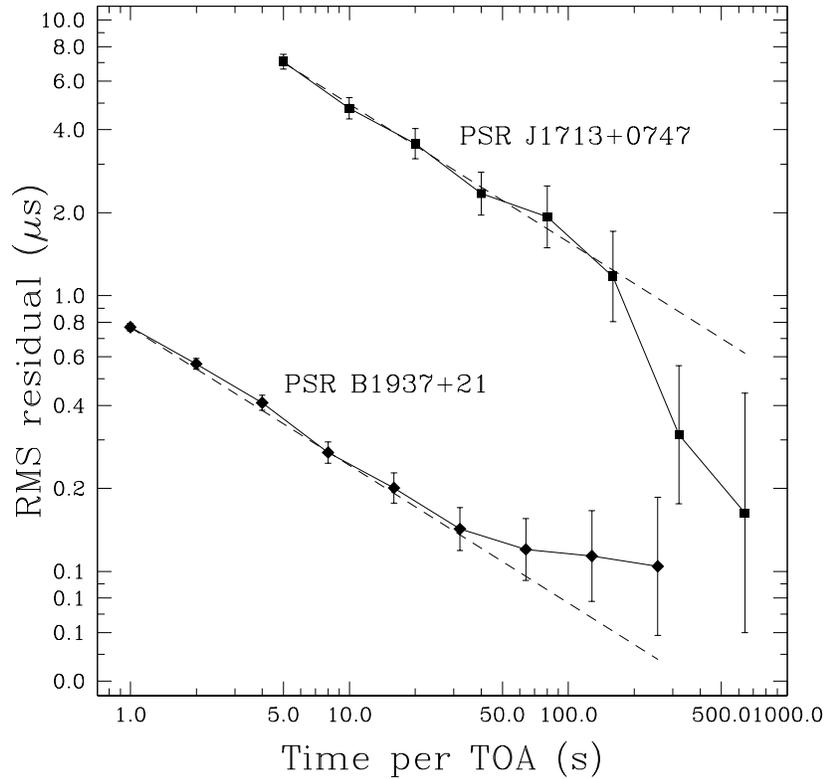}}
\end{picture}
\caption[] {Root-mean-square postfit timing
residuals for the millisecond pulsars PSRs~B1937+21 and J1713+0747, as
functions of the length of time $t$ used for the time-of-arrival
integrations.  The dashed lines indicate slopes corresponding to
$t^{-1/2}$.  See the text for a discussion of the departures from this
predicted slope.} \label{fig:stable}
\end{figure*}

If there are no systematic effects in a set of timing data, the
root-mean-square deviation of the times-of-arrival from the predicted
model should decrease as $t_{\rm int}^{-1/2}$, where $t_{\rm int}$ is
the integration time for the TOAs.  To test whether this holds true
for Mark~IV data, the millisecond pulsars PSRs~B1937+21 and J1713+0747
were observed for 30\,minutes with 10\,MHz bandwidth at the Arecibo
Observatory.  TOAs were then calculated for integration times ranging
from 1\,s to 640\,s and fitted to the pulsar timing model using the
{\sc tempo} program.  The rms postfit residuals were calculated for
each integration length; they are displayed as a function of $t_{\rm
int}$ in Figure~\ref{fig:stable}.  For PSR~J1713+0747, the residuals
closely follow the expected slope of $t_{\rm int}^{-1/2}$; the
apparent drop for large integration times can be explained by
small-number statistics.  For PSR~B1937+21, the rms residuals follow
the expected slope until leveling off at a deviation of approximately
100\,ns, indicating that systematic effects prevent greater timing
precision.

It is likely that the systematic effects limiting the timing precision
of PSR~B1937+21 are not due to the Mark~IV instrument but rather to
the variability of the pulses reaching the Earth.  Although
PSR~B1937+21, with a spin period of 1.56\,ms, is the fastest pulsar
known, and is indeed one of the most stable
\cite{ktr94}, its emission is subject to significant scattering by
the interstellar medium.  This process has the effect of convolving
the pulse profile with a variable exponential tail, resulting in
slightly changed observed profiles and hence an unavoidable
uncertainty in the TOA calculation.  The magnitude of this uncertainty
can be estimated by considering the pulsar scintillation due to
diffraction in the interstellar medium.  The size of the scintillation
features can be described by the decorrelation bandwidth, $\Delta
\nu$, and scintillation timescale, $\Delta t$, for a given observing
frequency.  At 1400\,MHz, these parameters for PSR~B1937+21 have been
found to be roughly $\Delta \nu = 0.83$\,MHz and $\Delta t = 400$\,s
\cite{rtd88}, leading to a scattering timescale of $\tau_{\rm s} =
1/(2\,\pi\,\Delta \nu) = 190$\,ns.  The portion of the timing
uncertainties due to scattering-induced profile changes should be
roughly equal to $\sigma_{\rm TOA} = \tau_{\rm s}\sqrt{\Delta \nu/B}$
\cite{cwd+90}, where $B$ is the observing bandwidth, yielding a
minimum uncertainty of 55\,ns for the 10\,MHz bandpass of the Mark~IV
observations.  While this is somewhat smaller than the observed lower
bound on the timing uncertainties, the order of magnitude is correct,
and indicates that the largest part of the remaining residuals is due
to scattering-induced variations of the pulse profile rather than to
instrumental systematics.  Furthermore, the timing analysis shows that
roughly halfway through the observation, the profile strength weakened
considerably due to scintillation for about 10 minutes, increasing
timing uncertainties for this period by roughly a factor of two.
Variable data quality will also prevent the residuals from integrating
down as predicted.  Though it is possible that there may be some
instrumental effects at the 50\,ns level, deriving, for instance, from
the 20\,MHz clock signal, it appears from these observations that the
overall timing properties of the \m4 instrument are satisfactory.

\subsection{Interference excision}\label{sec:rfi}

\begin{figure*}
\setlength{\unitlength}{1cm}
\begin{picture}(12,8)
\put(0,-4){\includegraphics{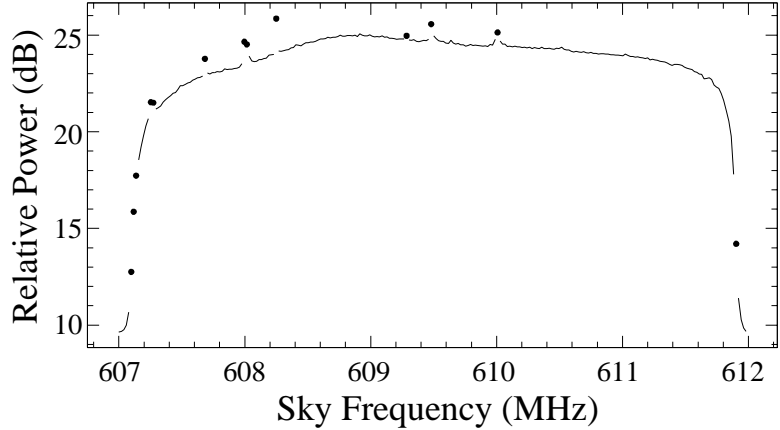}}
\end{picture}

\caption []
{An example of narrowband interference
  excision.  The connected lines are the acceptable parts of the
  spectrum; the points are those frequency bins that have been masked
  by the interference-hunting algorithm.}  \label{fig:nbzap}
\end{figure*}

An increasingly important problem facing radio astronomy is that of
radio frequency interference.  Communications satellites, television
stations and other transmitters render even the protected astronomy
bands in the radio spectrum vulnerable to noise.  Though negotiations
may bring about stronger protections, it is also useful to develop
data-acquisition instruments which are capable of mitigating its
effects.  Baseband recording systems permit interference to be
eliminated in software; the techniques discussed here could easily be
adapted for applications beyond pulsar observations.

The \m4 processing includes, optionally, two types of interference
excision: narrowband and broadband.  For narrowband excision, a short
(typically 256-point) power spectrum is computed for every tenth FFT
segment.  A simple algorithm searches through the spectrum,
calculating 5-point medians about every frequency bin and flagging
each point which differs from its corresponding median by more than
4\%, or which differs from both its nearest neighbours by more than
50\%.  Subsequently, points with both nearest neighbours or at least
three of four nearest neighbours flagged are also flagged.  A mask is
produced which is used to zero the contaminated frequencies in the
next set of data segments.  On average, 5 to 10$\%$ of frequency
points are zeroed in this fashion, with no apparent effect on the
resulting pulse shape.  Figure~\ref{fig:nbzap} shows a typical
spectrum and resulting mask.
 
\begin{figure*}
\setlength{\unitlength}{1cm}
\begin{picture}(12,10)
\put(-2,0){\includegraphics{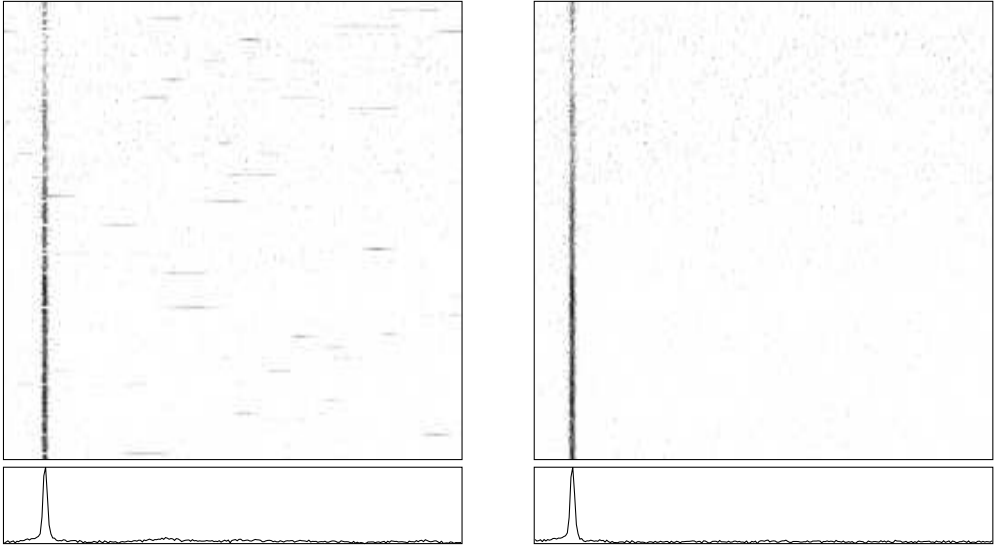}}
\end{picture}
\caption []
{An example of broadband interference excision.  The left panel shows
a grayscale plot of a contaminated 29-minute 610\,MHz observation of
PSR~B1534$+$12 taken with the 76\,m telescope at Jodrell Bank, U.K.
Each horizontal line represents ten seconds of data, and a cumulative
pulse profile is displayed at the bottom.  The broadband interference
spikes have been ``dedispersed'' by roughly 6$\%$ of the pulse period,
yielding the smeared-out bumps seen in the left-hand panel.  The right
panel presents the same observation, but with these broadband power
spikes eliminated during processing.} \label{fig:bbzap}
\end{figure*}

Broadband noise is eliminated by searching through the packed data
samples for power spikes greater than 30 times the root-mean-square
noise above the median.  If such spikes are found, the samples in
question are set to zero.  This algorithm is extremely effective at
eliminating broadband noise, again with no apparent effect on the
pulse profile shape.  ``Before'' and ``after'' grayscale plots of an
interference-contaminated observation are shown in
Figure~\ref{fig:bbzap}.  The benefits of interference excision may be
easily deduced from this figure: in some of the 10-second integrations
in the first analysis, the pulsar is drowned out by the interference
and does not appear at all, whereas in the second analysis it is
always present.  The baseline of the overall pulse profile is also
much improved.  This technique not only improves the signal-to-noise
ratio, but also eliminates a large source of systematic error in the
time-of-arrival fitting.

\section{CONCLUSION}\label{sec:concl}

The \m4 system provides an example of a 10\,MHz baseband recording
system designed for use in pulsar observations.  It meets and exceeds
the predicted improvements in timing precision over earlier
filterbank systems, and the analysis code contains features allowing
optimal profile-shape recovery for 2-bit quantization and the excision
of narrowband and broadband interference.  These improvements could
easily be generalized for application to other types of observations.

Future designs of pre-detection digital recorders will likely take
advantage of the availability of more powerful computers and faster
recording media to produce instruments with wider bandwidths.  The
performance-to-cost ratio of computers tends to increase
exponentially, doubling every 18 months.  Disk storage also becomes
faster and less expensive with time, while the cost of magnetic tapes
tends to remain more stable, thus the temptation is to migrate toward
wide-bandwidth systems with only disk storage.  Naturally, for an
instrument in which the data are processed directly from disk and
overwritten by subsequent observations, the reprocessing capability
provided by the more expensive and slower tape storage will be lost.
However, there are justifiable reasons for moving to such a scheme.
At frequencies above 1\,GHz, for instance, pulsar signals are often so
weak that more than 10\,MHz observing bandwidth is required to obtain
usable signal-to-noise ratios.  Further, although the precision of
measurements from strong, distant pulsars will be limited by
scattering, as discussed in \S\ref{sec:stable}, improved
signal-to-noise ratio is useful for nearly all other observations.
Finally, the inherent flexibility of baseband recording systems should
encourage the development of instruments which can be used for a
multitude of different types of centimetre-wavelength observations.

\section*{Acknowledgements} 
We thank Hal Taylor, Jay Shrauner and Phil Perillat for work in
designing the \m4 prototypes, Bob Wixted for helpful discussions, Stan
Chidzik for laying out and assembling the circuit boards, Mark
Krumholz, Christopher Scaffidi and Donald Priour for hardware work and
Walter Brisken for SAM-350 software testing.  The Arecibo Observatory,
a facility of the National Astronomy and Ionosphere Center, is
operated by Cornell University under a cooperative agreement with the
U.~S. National Science Foundation.  I. H. S. received support from an
NSERC 1967 fellowship. S. E. T. is an Alfred P. Sloan Research Fellow.
This work was supported by the NSF and the Seaver Institute.


\end{document}